\documentclass[aps,prb,twocolumn,superscriptaddress,amsmath,amssymb,longbibliography]{revtex4-2}
\usepackage{color,graphicx}
\usepackage{bm}
\usepackage{amsmath}
\usepackage{amssymb}
\usepackage{url}
\usepackage{epstopdf} 
\usepackage{wrapfig} 
\usepackage{lipsum}  

\usepackage[draft]{changes}
\begin{document}
\title{Prospects for a Solid-State Nuclear Clock}  

\author{Steven M. Girvin}
\email{steven.girvin@yale.edu}
\affiliation{Yale Quantum Institute, PO Box 208 334, 17 Hillhouse Ave, New Haven, CT 06520-8263, USA}
\author{Leo Radzihovsky}
\email{radzihov@colorado.edu}
\affiliation{Department of Physics and Center for Theory of Quantum
  Matter, University of Colorado, Boulder, CO 80309, USA}









\newcommand{\theauthors}{XXXXXX}
\newcommand{\thefirstauthors}{YYYYYY}

\begin{abstract}
  Motivated by recent experimental breakthroughs toward a realization
  of a solid-state Thorium-229 nuclear clock, we review the technology,
  basic physics motivation, and limitations of the present generation
  of atomic clocks.  We then discuss prospects for a new generation
  of clocks based on an anomalous low-energy 8.4 eV nuclear
  transition in Th-229, with an extremely long lifetime of 641 seconds
  when doped into CaF$_2$ crystals.  To realize such solid-state
  nuclear clocks one must confront basic nuclear, AMO, and solid state
  physics questions. Key challenges are understanding and minimizing the
  effects of inhomogeneous broadening, associated with strains and
  electric field gradients due to both the Th dopants and intrinsic crystal defects.
\end{abstract}

\maketitle

\maketitle

\section{Introduction and motivation}
An exciting new development in the quest for ever greater precision in measurement of time and frequency is the possibility of replacing current atomic clocks with a nuclear clock based on anomalously low-energy nuclear isomeric excitation lines in $^{229}$Th \cite{Peik_2003,Tiedau_2024,Elwell_2024,zhang_frequency_2024}. Typical electromagnetic excitations of nuclei correspond to x-ray and $\gamma$-ray energies ($10^4-10^6$ eV).  However, $^{229}$Th has long been expected to have a very low-energy excitation somewhere below $\sim 20$ eV.  After a decades-long search, recent breakthrough experiments \cite{Tiedau_2024,Elwell_2024} have finally located the extremely narrow optical resonance of the nucleus in the vacuum ultraviolet region near $8.4$ eV and directly excited it with laser light. While this excitation energy is an order of magnitude higher than that of optical clocks, it is still within the reach of modern (state of the art) lasers.  Indeed, a recent experiment \cite{zhang_frequency_2024} based on a precision optical comb (with coherent high-harmonic generation) has directly connected the frequency of UV laser light near $8.4$ eV to the JILA $^{87}$Sr clock and has determined the $^{229}$Th absolute excitation frequency (of the center of a particular $(I_\text{g}, 5/2) \rightarrow (I_\text{iso}, 3/2)$ spectral line within the multiplet of transitions) to be $2,020,407,298.727(4)$ MHz.

As we will discuss below, this nuclear isomeric transition has numerous advantages for use in precision timekeeping and quantum sensing, provided that complications associated with having the nuclei in a solid-state host can be overcome:
\begin{itemize}
    \item The clock `ticks' faster because the isomeric transition frequency is high relative to traditional microwave and optical clocks. 
    \item The transition matrix element is extremely weak, leading to a long lifetime for the excited state to spontaneously emit a photon (in free space) of $T_1\sim 2,000 s$.  [The index of refraction in the CaF$_2$ host crystal lowers the speed of light and thus increases the density of states for photons, reducing the measured lifetime \cite{zhang_frequency_2024} to $\sim 641\pm 4$ s.]
    One thus expects an extremely narrow (a few millihertz out of $\sim 2$ petahertz) {\em intrinsic} spectral width, 
    which would yield extremely good frequency resolution.  
    \item The nucleus is very small in size, and its energy levels are only weakly perturbed by stray electric fields and other spurious effects that lead to systematic errors in traditional atomic clocks.
    \item Small numbers of Th$^{3+}$ ions can be stored and manipulated in an ion trap (using now-standard technology), providing a very clean environment to minimize spurious effects.
    \item Huge numbers of Th$^{4+}$ ions can be incorporated into solid-state crystals, offering potentially enormous increases in signal-to-noise ratios for precision measurements (if systematic errors due to spurious solid-state effects, discussed below, can be controlled).
\end{itemize} 

It took many decades of effort to eliminate the sources of systematic errors in optical clocks \cite{Ludlow_2015_RevModPhys.87.63,Aeppli_2024_optical_clock}. If the same can be achieved for a nuclear clock, we can anticipate breakthroughs in GPS, metrology, and the exploration of fundamental physics through tabletop experiments.  Extreme precision in the measurement of time and frequency could potentially one day be used for the detection of primordial gravitational waves and perhaps even allow us to explore quantum effects in gravity.

The exceptionally low nuclear isomeric excitation energy of $\sim 8.4$ eV in $^{229}$Th is the result of an accidental cancellation between two very large energy scales (both of which are on the order of $\sim 50$ keV): the decrease in Coulomb energy and the increase in strong-force energy as the nucleons rearrange upon ground-to-isomer state excitation.  The large values of these energies and their near cancelation make the $^{229}$Th nucleus an excellent testing ground to look for time variation in the fine structure constant (which would affect the Coulomb energy) or to search for interactions with dark matter (which might affect the strong-force energy). Ref.\ \cite{beeks_fine-structure_2025} shows that $^{229}$Th has sensitivities to these `beyond Standard Model' effects, which are enhanced by four orders of magnitude due to the large but canceling energy scales involved in the nuclear excitation.  

\section{Brief Background on Atomic Clocks}
Atomic clocks suffer from three classes of limitations: 
\begin{itemize}
    \item The first is the finite lifetime $T_1$ of the excited atomic state due to spontaneous emission of photons.  The time-frequency uncertainty relation guarantees that this produces a finite linewidth for the transition that is not less than $\delta\omega= \frac{1}{2T_1}$. This can be remedied by selecting a quasi-forbidden (i.e., not electric-dipole-allowed) atomic transition, which thus has a long $T_1$.
    \item  The second is $\sqrt{N}$ shot noise associated with the finite number (typically $N\sim 10^4-10^5$) of atoms that are being interrogated.
     \item The third class of limitations consists of a myriad of subtle systematic perturbations resulting from stray electric and magnetic fields, black-body radiation, and many-body effects associated with weak interactions among the atoms (e.g., dipolar coupling). Such effects can cause dephasing of the clock transition (homogeneous broadening) as well as inhomogeneous broadening because of frequency shifts that vary with the positions of individual atoms. It has taken a substantial effort over the last decade to find ways to suppress systematic errors associated with many-body effects in optical atomic clocks so that the number of atoms can be scaled up to reduce the shot noise.  Such many-body effects are expected to be smaller in nuclear clocks due to the small size of the nucleus.
\end{itemize}
   
The microwave excitation of Cs, associated with the hyperfine interaction, has long been used to construct extremely precise atomic clocks ($\sim 10^{-16}$ at 1 second, which corresponds to losing 1 second in 100 million years!) and is currently used to define the second in the SI system of units.  The lifetime of the Cs excited hyperfine state is effectively infinite in practical experiments; therefore, the frequency uncertainty is limited by the interrogation time ($\sim 1$~s in `fountain' clocks). Modern atomic clocks \cite{Ludlow_2015_RevModPhys.87.63,Aeppli_2024_optical_clock} based on optical transitions in atoms held (for long periods) in optical lattices have interrogation times limited by the spontaneous emission lifetime of the excited state. Optical transitions possess much higher frequencies but typically have much shorter spontaneous emission lifetimes of tens of nanoseconds.  However, by using states with quasi-forbidden optical transitions, long lifetimes ($T_1\sim 10-100$ s) can be achieved. The combination of a high oscillation frequency and a long lifetime provides optical clocks with a significant natural advantage in timekeeping.  $^{87}$Sr optical clocks \cite{Aeppli_2024_optical_clock} are approaching precisions of $10^{-20}$ and are so precise that the gravitational redshift associated with a height difference as small as $\sim 1$ mm can be measured \cite{bothwell_resolving_2022}.  As discussed above, the nuclear $^{229}$Th transition near $8.4$ eV is another order of magnitude higher in frequency and has an even longer lifetime of $T_1\sim 641\pm 4$ s, thus holding great promise for next generation clocks.

 Embedding $^{229}$Th in a solid-state host (such as CaF$_2$ crystal) offers the possibility of increasing $N$ by as much as 10 to 15 orders of magnitude compared to typical atomic clocks, which would dramatically improve the signal to noise ratio (which scales as $\sqrt{N}$).
A key challenge is to characterize, understand, and reduce significant systematic errors and inhomogeneous broadening of the nuclear clock transition that the solid-state host introduces.  The radius of the charge distribution of a nucleus is about five orders of magnitude smaller than that of the electron cloud in an atom, making nuclear excitation frequencies much less sensitive to external perturbations than electronic excitation frequencies. Nevertheless, at the extreme levels of precision needed to outperform atomic clocks, even tiny solid-state effects at the $10^{-11}$ level can be quite serious, as we shall describe in more detail below. 

In order to understand the systematic errors associated with solid-state effects, we first need to understand the nuclear physics of $^{229}$Th, which we address in the next section.

\section{The nuclear physics of $^{229}$Th}
The long lifetime ($\sim 641\pm 4$ s in the CaF$_2$ host material) for the spontaneous emission of a photon by the isomeric state of $^{229}$Th is a result of the tiny size of the nucleus and the fact that the transition is electric-dipole forbidden.  The transition is of  mixed magnetic dipole (M1) and electric quadrupole (E2) character \cite{zhang_frequency_2024}, resulting in a very weak matrix element for spontaneous photon emission, which is beneficial for minimizing the linewidth; however, it also makes the transition very difficult to drive with a laser.

A key feature to understand is that it is essential to work with highly ionized Th atoms (charge states 3+ or 4+) because the electron ionization energy of the lower charge states is less than $8.4$ eV.  Otherwise, the small overlap of the valence electrons with the nucleus induces a direct relaxation of the isomeric state to the ground state, with simultaneous ejection of a valence electron from the atom.  This decay channel in non-ionized Th then dramatically shortens the excited state lifetime from $\sim 641$ s down to $\approx 7\pm 1\, \mu$s \cite{Seiferle_2017}.  

Early theoretical \cite{wang2025directnuclearlevelqubitsusing,PhysRevLett.130.103201} and experimental \cite{Yamaguchi_2024} investigations are underway  to study Th$^{3+}$ ions held in an ion-trap. However, to  date, experiments have been conducted mainly with thin films or bulk crystals of CaF$_2$ doped with Th$^{4+}$ ions.  CaF$_2$ is a widely used optical material that has a band gap larger than $8.4$ eV (the indirect band gap is 11.8 eV and the direct band gap is 12.1 eV), making it transparent to the VUV light used to excite the nuclei and to the light spontaneously emitted by the nuclei (which is used to detect successful excitation).  This also makes direct isomeric nuclear decay through electron excitation into the conduction band energetically forbidden, thereby maintaining the intrinsic long $T_1$ time.


The ground state of $^{229}$Th has a nuclear spin quantum number $I_\mathrm{g}=\frac{5}{2}$, while the excited isomeric state has spin $I_\mathrm{iso}=\frac{3}{2}$. Both states of the nucleus are non-spherical and have the shape of a symmetric top (i.e., a spheroid), implying that both nuclear states carry an electric quadrupole moment that can couple to electric field gradients at the position of the nucleus (see the Appendix for a detailed derivation).  The degree of shape distortion and the charge radius are slightly different for the two $I$ states \cite{zhang_frequency_2024}. The quadrupolar coupling splits the two states into multiplets, with energy splittings on the order of 100 MHz. This can be advantageous for combined optical, NQR, and NMR resonances for the purposes of quantum control and readout of the nuclear qubit.  It is important to develop readout methods that are much faster than simply monitoring the fluorescence emission from the Th ions, which is very slow due to the long $T_1$ time scale.

\section{Solid State Effects}

Complex nuclear physics controls the $\Delta E_I = 8.4$ eV splitting between thorium's  6-fold degenerate nuclear ground state characterized by spin quantum number $I_\mathrm{g}=\frac{5}{2}$ and the 4-fold degenerate excited isomeric state with spin $I_\mathrm{iso}=\frac{3}{2}$. Once thorium is doped into the CaF$_2$ crystal, the nucleus is exposed to numerous perturbations controlled by solid-state physics. As was revealed in the experiments\cite{zhang_frequency_2024,Jun_Temperature_2025}, which we highlight in the following section, the most important of these arise from the local electric field gradients interacting with the nuclear quadrupolar moment. As derived in the Appendix, this is characterized by the quadrupolar coupling Hamiltonian, $\hat H_Q = e \hat Q_{\mu\nu} V_{\mu\nu}$, which in the crystal's frame is given by,
\begin{align}
    \hat H_Q=\frac{e Q_I V_{zz}}{4I(2I-1)}[3I_z^2-I(I+1)+\eta(\hat I_x^2-\hat I_y^2)].
\label{Hquad}
\end{align}
Here, $Q_I$ is the electric quadrupole moment, $I$ is the nuclear spin quantum number, $-V_{zz}$ is the $zz$ component of the electric field gradient tensor, and $\eta = \frac{(V_{xx}-V_{yy})}{V_{zz}}$ is an asymmetry parameter for the electric field gradient in the plane transverse to $z$, quantifying the breaking of tetragonal symmetry. To include the ground-isomer $\Delta E_I = 8.4$ eV splitting, the full Hamiltonian is $\hat H_I^\text{Th229}
=E_I + \hat H_Q$. The quadrupolar coupling perturbation $\hat H_Q$ is straightforwardly diagonalized\cite{Zory1965,Gerdau1969} (noting that $\eta$ couples to $\hat I_+^2 + \hat I_-^2$, with the selection rule $\Delta I_z =\pm 2$), generically splitting the ground $I_\mathrm{g}=5/2$ and isomer $I_\mathrm{iso} = 3/2$ nuclear states into 3 doublet lines, $|I_z| = 5/2, 3/2, 1/2$ and 2 doublet lines, $|I_z| = 3/2, 1/2$. 

The detailed nature of the above splitting depends on the location of Th within the CaF$_2$ unit cell, and the configuration of the screening charges. For Th$^{4+}$ occupying the body-center,
the configuration has cubic symmetry and hence a vanishing electric field gradient $V_{zz} = \eta = 0$. This may explain the observation of an apparently unsplit line reported in low-resolution measurements by Hiraki et al.\ \cite{hiraki_2025}. For all other configurations, we expect non-zero quadrupolar splitting and these authors observed other lines corresponding to substantial quadrupolar splittings.
Recent high resolution measurements of the quadrupolar splitting \cite{zhang_frequency_2024} have obtained $V_{zz}\approx 109.1(7) \mathrm{\, V}/\mathrm{\AA}^2$.  This paper used a different type of sample preparation compared to Hiraki et al., and did not see evidence of an unsplit line from the putative sites with cubic symmetry.

\section{Discussion of recent precision measurement results}

Having outlined the motivation, background, and challenges associated with the new generation of nuclear clocks based on $^{229}$Th transitions, we now discuss in more detail the breakthroughs made in three recent papers. 

\subsection{Frequency ratio of the $^{229}$Th nuclear isomeric transition and the 87Sr atomic clock, Zhang et al. 2024}

The paper by Zhang et al.\ \cite{zhang_frequency_2024} reports the first direct frequency measurement of the $^{229}$Th nuclear clock transition, connecting it with very high precision to the $^{87}$Sr optical atomic clock and reporting the $5$  quadrupole splitting transitions, as well as its effective average transition frequency (i.e., corresponding to the $I_g = 5/2\rightarrow I_\text{iso}=3/2$ transition free of the electrical field gradient) 
$\nu_\mathrm{Th}=2,020,407,384.335(2)$ MHz ($8.3557335(8)$ eV) and $\nu_\mathrm{Th}/\nu_\mathrm{Sr}=4.707072615078(5)$. The measurement improved the precision of the $^{229}$Th transition frequency by six orders of magnitude compared to earlier estimates. This realizes the first direct frequency ratio between a nuclear and an electronic transition, marking a milestone toward a nuclear optical clock and launching a new era of timekeeping and table-top tests of fundamental physics. 

The authors used an optical frequency comb locked to the ultra-stable JILA strontium optical-lattice clock, followed by coherent seventh harmonic generation,  to directly excite the $^{229}$Th transition in a  CaF$_2$ crystal. With this technique, the experiment resolved five distinct hyperfine lines of quadrupolar splitting of the nuclear ground state to isomer excitation at 8.4 eV (149 nm). The splitting is associated with the interaction of the thorium nuclear electric quadrupole $Q_{ij}$ with the local electric field gradients $V_{ij}=-\partial_i E_j$ of the crystal lattice. The corresponding quadrupolar Hamiltonian $\hat H = e \hat Q_{ij} V_{ij}$ gives 6 levels (3 doublets, $|I_z| = 5/2, 3/2, 1/2$) for the ground $I_\mathrm{g}=5/2$ nuclear state and 4 levels for the excited isomer $I_\mathrm{iso} = 3/2$ nuclear state (2 doublets, $|I_z| = 3/2, 1/2$).

From these splittings, the authors extracted several key parameters, including (i) the ratio of the spectroscopic quadrupole moments between the isomeric and ground states, $Q_\mathrm{iso}/Q_\mathrm{g} = 0.57003(1)$, (ii) the average  field gradient in the crystal of $V_{zz}=109.1(7) V/{\mathrm{\AA}}^2$, and (iii) the excited-state lifetime of $T_1=641\pm 4$ seconds, which is consistent with previous findings. Notably, the spectroscopic line widths were found to be $\sim 300$ kHz (reduced in new samples more recently to 20 kHz \cite{ooi2025frequencyreproducibilitysolidstateth229}), some 8 orders of magnitude wider than the expected intrinsic width $1/(2T_1)$.  This excess linewidth is thought to be associated with inhomogeneous broadening due to impurity-induced local crystal strains and field gradients, which we discuss below. 
The authors found that the line centers remained stable over two weeks of data collection, demonstrating 
a remarkable environmental stability, which is a key requirement for a robust clock.  

\subsection{Temperature Sensitivity of Thorium-${229}$ Solid-State Nuclear Clock, Higgins et al.\, {\sl Phys.\ Rev.\ Lett.\ }{\bf 134} (11): 113801 (2025) }

In this paper \cite{Jun_Temperature_2025}, the authors make precise measurements of the temperature dependence of the transition frequency of the nuclear transition in a CaF$_2$ host crystal.  Importantly, they find that one of the lines in the quadrupolar multiplet (the $I_z =\pm 5/2 \rightarrow  I_z=\pm 3/2$ transition) exhibits weak temperature dependence, meaning that with adequate temperature control, temperature fluctuations will not limit the clock stability. For example, a $5\,\mu$K control over temperature would correspond to $10^{-18}$ fractional frequency uncertainty.

We can understand this effect from a number of subtle contributions to the temperature dependence of the transition frequency.  Naively, one might expect thermally excited phonons to have a significant effect. However, the phonons that modulate the atomic positions (and hence the quadrupolar splittings) are very high in frequency compared to the narrow linewidth and are expected to almost completely average out. Residual anharmonicities (phonon-phonon interactions) lead to the thermal expansion of the lattice with increasing temperature, which lowers the electric field gradients and, hence, the quadrupolar splitting. The associated change in the splitting causes some of the lines to move up and others to move down in frequency.  At the same time, the lattice expansion (and possibly other effects not yet determined) slightly reduces the electron density $\rho(0)$ at the position of the $^{229}$Th nucleus.   As mentioned in the Appendix, this electron density contributes to the electric field gradient through the Poisson equation, $\nabla^2 V=-\frac{1}{\epsilon_0}\rho(0)$. This, in turn, creates a monopole shift in the line center towards higher frequency.  It happens that the two effects approximately cancel each other for the $I_z =\pm 5/2 \rightarrow  I_z=\pm 3/2$ transition, which results in the observed  very weak temperature dependence for the transition frequency.   

\subsection{Frequency reproducibility of solid-state Th-229 nuclear clocks, Ooi et al., arXiv:2507.01180 [Nature (in press)]}
More detailed temperature measurements by the same authors in this most recent work  \cite{ooi2025frequencyreproducibilitysolidstateth229} have actually found a temperature at which the transition frequency is a minimum, meaning that it is only second-order sensitive to temperature fluctuations.  Operating at this temperature ($\sim 190$ K) greatly reduces demands on the temperature control system for a given frequency stability.

By comparing different samples, the authors demonstrate that the line width is linear in the concentration of $^{229}$Th, indicating that the doping itself contributes to the mechanical strains and charge inhomogeneities in the host crystal.  Intrinsic strains are  the likely source of residual inhomogeneous line broadening of the order $20$ kHz observed in the limit of low doping \cite{ooi2025frequencyreproducibilitysolidstateth229}. Although this is only one part in $10^{11}$ of the optical transition frequency of the nucleus, it is vastly larger than the ultimate linewidth limit set by the finite lifetime of the isomeric state $\frac{1}{2T_1}\sim 10^{-3}$ Hz. 

This work looked at the stability of the line center frequency relative to the JILA $^{87}$Sr clock over a much longer period (nearly one year) and found no systematic drift, which indeed bodes well for the future of nuclear clocks.

\section{Discussion and Open Questions}
As described above, based on the 641 second lifetime \cite{zhang_frequency_2024}, the {\em intrinsic} 8.4 eV nuclear spectral lines in thorium are expected to be extremely narrow, on the order of a couple of milli-Hertz. Yet, when doped into a CaF$_2$ crystal, these lines exhibit a width of tens of kilo-Hertz. A detailed understanding of this 8 orders of magnitude of inhomogeneous broadening (relative to the intrinsic linewidth)-- spatially random shifts in quadrupolar splitting frequencies $\delta\omega_n$ -- and its potential mitigation remain key challenges to the realization of a solid-state nuclear clock.

As is typical for all realistic crystals, CaF$_2$ is characterized by many sources of heterogeneity, including vacancies and interstitials, dislocations and grain-boundaries, leading to random strain fields.  In addition, there are random electric fields (and hence field gradients) associated with charged impurities. These lead to a random distribution of quadrupolar splittings that are almost certainly responsible for the dominant contributions to the aforementioned inhomogeneous broadening of the $^{229}$Th line shape, which appears to be Lorentzian, unusual for inhomogeneous broadening, and with a width that grows linearly with Th dopant concentration \cite{ooi2025frequencyreproducibilitysolidstateth229}.  Stoneham \cite{Stoneham_1969} presents an argument for the origin of the wings of the frequency distribution that fall off slowly ($\sim (\delta\omega)^{-2}$ as in a Lorentzian) due to the $1/r^3$ power law dependence of the electric field gradient on the distance of an impurity from the Th nucleus.  Rare events in which there is an impurity close to the Th nucleus account for the long tail in the frequency distribution.
In addition to quadrupolar splitting, there is, as mentioned above, a small monopolar contribution to the broadening of the transition frequency associated with inhomogeneities in the local electron density at the position of the $^{229}$Th ions \cite{Jun_Temperature_2025}.

These disorder effects are exacerbated by the fact that the $^{229}$Th ions must be doped into the CaF$_2$ crystal.  There is little firm understanding at this point about the sites at which the Th ions tend to sit and what the configuration of compensating charges is in the neighborhood of these sites.  Hiraki et al.\ \cite{hiraki_2025} have suggested, based on DFT calculations and the observed large quadrupolar splitting, that an important charge configuration is pairs of Th$^{4+}$ ions relatively close together.  Much more work is needed on first-principles calculations to help us understand the most probable atom configurations and the details of the inhomogeneities that determine the linewidths.  

A complicating factor is that the electric field gradients at the $^{229}$Th positions are not easy to compute.  Simple point-charge models for an ionic crystal with defects do not necessarily work because of complications from Sternheimer anti-screening \cite{Sternheimer_1951,Sternheimer_1966,Slichter_NMR}: the electric fields polarize the electron cloud around the Th nucleus in a way that can dramatically enhance the quadrupolar splitting by 1-2 orders of magnitude. These effects will need to be carefully taken into account in any first-principles calculations. 
Another crucial task for experimenters is to understand whether there are different sample preparation or annealing methods that can sharply reduce the inhomogeneously broadened line widths by many orders of magnitude.

The $^{229}$Th nuclear transition is, by ordinary standards, extremely insensitive to external perturbations.  However, by today's exacting standards for atomic clocks, much work remains to be done to assemble a large collection of $^{229}$Th$^{4+}$ ions within a very homogeneous environment to realize the dream of a compact highly precise nuclear clock.  If this challenge can be overcome (as it was in the history of atomic clocks) there will be great opportunities for both fundamental research and practical applications.

{\it Acknowledgments}. We thank members of the Ye nuclear clock group,
Emil Pellett, and especially Jun Ye and Tian Ooi for discussions. SMG thanks Vidvuds Ozolins for discussions and 
acknowledges support from JILA as a Visiting Fellow and support from Yale University.
This work was supported by the Simons Investigator Award
to LR from the James Simons Foundation.

\section{Appendix: Derivation of the Quadrupolar Coupling}

Any nucleus having a spin greater than $1/2$ is permitted (but not required) by the Wigner-Eckhart theorem to have an electric quadrupole moment. For the case of a symmetric top, the \emph{intrinsic} quadrupole moment of the charge distribution is defined by a single scalar number computed in the body frame of the nucleus (i.e., with the symmetry axis of the nucleus aligned with the $z$ axis of the coordinate system) by the integral 
\begin{align}
    eQ=\int d^3r\, \rho(\vec r) (3z^2-r^2).
\end{align}
The ground state of $^{229}$Th has a nuclear spin $I=5/2$ and a \emph{spectroscopic} quadrupole moment $eQ_0\approx 3.11$e-barn, while the excited isomer state has a spin $I=3/2$ and a quadrupole moment $eQ_1\approx 1.74$e-barn (where $1 \mathrm{\ barn}\equiv 10^{-28}\mathrm{m}^2= 100 \mathrm{fm}^2$).

Such nuclear quadrupolar moments couple to electric-field gradients in solid-state hosts and have long been observed through nuclear quadrupolar resonance (NQR) and NMR techniques.  The quadrupolar contribution to the electrostatic potential energy of the nucleus in the presence of a local potential $V(\vec r)$ in the laboratory frame is given by
\begin{align}
    U=\frac{1}{2}eq_{\mu\nu}V_{\mu\nu},
\end{align}
where summation over repeated Greek indices is assumed, and 
\begin{align}
    V_{\mu\nu}=\partial_\mu\partial_\nu V(\vec r)|_{\vec r=\vec 0}.
\end{align}
The nuclear electric quadrupole tensor in the laboratory frame (not the body frame) is given by
\begin{align}
    eq_{\mu\nu}=\int d^3r\, r_\mu r_\nu \,\rho(\vec r).
\end{align}
If we work in a laboratory frame in which $V_{\mu\nu}$ is diagonal, then
\begin{align}
    U=\frac{e}{2}[V_{zz}q_{zz}+V_{xx}q_{xx}+V_{yy}q_{yy}].
\end{align}
Assuming (though this is not quite correct) that there are no other charges present that overlap the nuclear charge, the Laplace equation gives $V_{\mu\mu}=0$ (summation over $\mu$ implied), and we can write
\begin{align}
    U=\frac{e}{4}V_{zz}[3q_{zz}-\{q_{xx}+q_{yy}+q_{zz}\}+\eta(q_{xx}-q_{yy})],
\end{align}
where 
\begin{align}
    \eta = \frac{(V_{xx}-V_{yy})}{V_{zz}}.
\end{align}
Including the electron charge density $\rho(0)$ at the position of the nucleus, the Laplace equation is replaced by the Poisson equation $V_{\mu\mu}=-\frac{1}{\epsilon_0}\rho(0)$.  This introduces a small (but significant) monopole shift in the transition frequency, which we will neglect here.

The Wigner-Eckart theorem tells us that, within a single spin multiplet, the quantum operator representing the quadrupole moment can always be written solely in terms of the spin operators
\begin{align}
    eq_{\mu\nu}=\frac{eC(I)}{2} [I_\mu I_\nu+I_\nu I_\mu],
\end{align}
where symmetrization is included to enforce hermiticity, and $C(I)$ is a constant that depends only on the total spin quantum number. Using this theorem, we can write the quantum operator for the quadrupolar term in the Hamiltonian as
\begin{align}
    U=\frac{eC(I)}{4}V_{zz}[3I_z^2-I(I+1)+\eta (I_x^2-I_y^2)].
\end{align}

It can be shown [Slichter \cite{Slichter_NMR}, Eq.~(10.50)] that the constant $C(I)$ is related to the 
body-frame charge distribution in such a way that, in the maximum weight state ($I_z=I$), 
\begin{align}
    Q_I&=C(I) [3I_z^2 -I(I+1)],\\
    C(I)&=\frac{Q_I}{I(2I-1)}.
\end{align}
Hence, we finally arrive at the following convenient form:
\begin{align}
    U=\frac{eQ_IV_{zz}}{4I(2I-1)}[3I_z^2-I(I+1)+\eta(I_x^2-I_y^2)].
\end{align}
The diagonalization of this Hamiltonian gives the $2I+1$ eigenvalues of the spin multiplet due to the quadrupolar interaction, clearly doubly-degenerate with respect to time-reverse doublets $\pm I_z$.  Since the Hamiltonian is traceless, there is no mean shift of the levels relative to the unperturbed Hamiltonian. (Note that this is not true when the Laplace equation is invalid due to electronic charge overlapping the nucleus.  Also note that the body frame quadrupole moment $Q_I$ depends on the isomeric state and hence on $I$.) 

To get a sense of the scale of the quadrupole splitting, let us consider the Th nucleus at a site with cubic symmetry (hence no quadrupolar splitting).  We perturb this with an additional charge of strength $Ze$ at a distance $R$ from the Th nucleus.  If we neglect dielectric screening (probably valid for small $R$ but not for large $R$), the quadrupolar energy scale is 
\begin{align}
    U_0(R) = \frac{Q_IZe^2}{2(4\pi\epsilon_0)R^3}. 
\end{align}
Taking $Z=4$ and the radius to be one lattice constant, $R\approx 5\mathrm{\AA}$, yields
\begin{align}
    V_{zz}=\frac{2Ze}{4\pi\epsilon_0 R^3}\sim 1\mathrm{\ V}/\mathrm{\AA}^2,
\end{align}
which is two orders of magnitude smaller than the experimental value (inferred from the measured quadrupolar splitting). However, it is quite possible that Sternheimer anti-screening \cite{Sternheimer_1951,Sternheimer_1966,Slichter_NMR} (associated with the polarization of core electrons near the nucleus) can enhance the electric field gradients by as much as two orders of magnitude. Hiraki et al.\ \cite{hiraki_2025} used DFT calculations to find a nuclear configuration yielding 
$V_{zz}\approx 95 \mathrm{\, V}/\mathrm{\AA}^2$, with an asymmetry
parameter of $\eta\approx 0.6$.  A high-precision optical comb
measurement by Zhang et al.\ \cite{zhang_frequency_2024} found
$V_{zz}\approx 109.1(7) \mathrm{\, V}/\mathrm{\AA}^2$, with an
asymmetry parameter of $\eta\approx 0.59163(5)$.  Hiraki et al.\
\cite{hiraki_2025} have presented electronic structure calculations
and experimental evidence for the existence of pairs of closely spaced
$^{229}$Th defects that are consistent with the observed value of
$V_{zz}$. It seems surprising that two $Z=4$ defects would sit so
close together, but Hiraki et al.\ show that, in addition to
approximately giving the correct $V_{zz}$, this pair of defects would
also approximately produce the observed value of the asymmetry
parameter $\eta$ inferred from the measured quadrupole splitting.

\bibliographystyle{unsrt}
\bibliography{Nuclear_Clock_Bibliography}

\end{document}